\pdfoutput=1
\documentclass[aps,prb,twocolumn,superscriptaddress,floatfix,10pt]{revtex4-2}
\usepackage[version=4]{mhchem}
\usepackage{graphicx}
\usepackage{amsmath}
\usepackage{float}
\usepackage{xcolor}
\definecolor{Rev}{rgb}{0.75, 0.00, 0.00}

\begin{document}


\title{Microscopic correlation between magnetostriction and magnetic damping}


\author{Ivan Kurniawan}
\email{kurniawan.ivan@nims.go.jp} 
\affiliation{Research Center for Magnetic and Spintronic Materials, National Institute for Materials Science (NIMS), Tsukuba 305-0047, Japan}
\author{Keita Ito}
\affiliation{Institute for Materials Research, Tohoku University, Sendai 980-8577, Japan}
\author{Takeshi Seki}
\affiliation{Institute for Materials Research, Tohoku University, Sendai 980-8577, Japan}
\affiliation{Center for Science and Innovation in Spintronics, Tohoku University, Sendai 980-8577, Japan}
\author{Keisuke Masuda}
\affiliation{Research Center for Magnetic and Spintronic Materials, National Institute for Materials Science (NIMS), Tsukuba 305-0047, Japan}
\author{Yoshio Miura}
\affiliation{Research Center for Magnetic and Spintronic Materials, National Institute for Materials Science (NIMS), Tsukuba 305-0047, Japan}
\affiliation{Faculty of Electrical Engineering and Electronics, Kyoto Institute of Technology, Matsugasaki, Sakyo-ku, Kyoto, 606-8585, Japan}

\begin{abstract}
Although the relationship between magnetostriction and magnetic damping is often described phenomenologically, their intrinsic connection remains unclear. In this study, we demonstrate that the magnitude of magnetic damping depends on the sign of magnetostriction in \ce{(Fe_{1-x}Co_{x})_{4}N} and \ce{Ni_{1-y}Co_{y}} alloys across various compositions, consistent with experimental observations. This behavior is attributed to strain-induced changes in exchange splitting, which shift the minority spin density of states near the Fermi level, thereby affecting both magnetostriction and damping through spin-conserving transitions. Additionally, the presence of locally degenerate orbitals plays a crucial role in determining magnetostriction. These findings suggest that magnetization dynamics and magnetostriction can be intrinsically controlled, facilitating the design of magnetic materials for applications such as flexible spintronics.
\end{abstract}

\maketitle


The first observation of the spatial deformation of magnetic materials under an applied magnetic field was reported by Joule for iron in 1842 \cite{joule1842}, later termed magnetostriction ($\lambda$), which is defined as a change in length ($\delta l/l$) \textcolor{black}{between the demagnetized state and the magnetically saturated state along the magnetic field direction}. At that time, Joule found that the elongation in soft iron bars is more prominent than in hardened steel \cite{joule1842}, implying that magnetostriction is inversely proportional to the elastic constants. It was later understood that contraction (negative magnetostriction) is exhibited in nickel \cite{bidwell1890}; hence, magnetostriction is not solely determined by the elastic constant but is also affected by the strain dependence of magnetic anisotropy energy (MAE) (which we refer to here simply as magnetoelasticity). When magnetoelasticity is nonzero, the preferred magnetic direction is changed by the strain; hence, the crystal structure will spontaneously deform in order to lower the total energy, which could result in elongation or contraction \cite{kittel1949}. Magnetostriction is expressed as the ratio of the magnetoelasticity to the elastic constant \cite{kittel1949}. Since the elastic constant of typical magnetic materials is generally expected to have a positive sign, magnetoelasticity predominantly determines the sign and magnitude of magnetostriction over compositional change. Subsequently, first-principles calculations were introduced by Wu \textit{et al.} \cite{wu1997}, providing intrinsic insights into the mechanisms of magnetostriction. This sparked significant interest, especially after the discovery of giant magnetostrictive materials like Terfenol \cite{clark1972}.

\begin{figure}[h]
\includegraphics[width=\columnwidth]{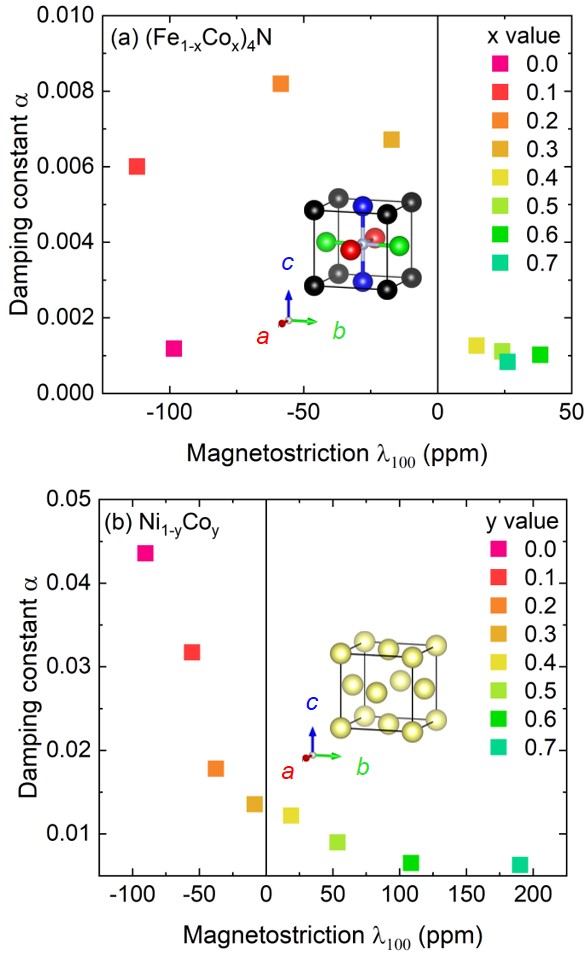}%
\caption{\label{fig1} \textcolor{black}{Calculated magnetostriction ($\lambda_{100}$) and damping constant ($\alpha$) for (a) \ce{(Fe_{1-x}Co_{x})_{4}N} and (b) \ce{Ni_{1-y}Co_{y}}. Crystal structure of (a) cubic \ce{Fe4N} and (b) fcc-Ni, showing sites of Fe1 (black), Fe2 (blue), Fe3 (green), Fe4 (red), N (grey), and Ni (yellow) with the corresponding coordinate system}}
\end{figure}

Around the same time as the establishment of first-principles studies on magnetostriction, a phenomenological model was proposed to explain the role of magnetostriction in the magnetic damping constant ($\alpha$) \cite{suhl1998}. This model introduced interrelated equations of motion for both strain and magnetization, resulting in temporal relaxation expressed as magnetic damping, which is proportional to the square of magnetostriction. This interpretation was further supported by other phenomenological models \cite{vittoria2010,rossi2005}. Experimentally, Heusler alloys \cite{chumak2021} demonstrated this proportional relationship. Meanwhile, variations in the composition of \ce{Ni_{x}Fe_{1-x}} revealed two distinct regions: negative (positive) magnetostriction was associated with a large (small) damping constant, though the underlying mechanism remains unknown \cite{endo2018}. It is significant to note that both magnetoelasticity and magnetic damping originate from spin-orbit interaction (SOI) \cite{wu1997,kambersky1976}. Moreover, the relationship between magnetostriction and magnetic damping is particularly valuable for spintronics applications, as it enables simultaneous optimization of energy efficiency and multidirectional sensing, which are critical for flexible spintronics technologies \cite{ota2018}.

Since SOI links magnetostriction and damping, a conventional approach to control both would involve tuning SOI strength, $\xi$, as expressed in the spin-orbit Hamiltonian $H_{\mathrm{SO}}=\sum_{I}\xi_{I}\vec{L}\cdot \vec{S}$, where $\vec{L}=(L_{x},L_{y},L_{z})$ and $\vec{S}=(S_{x},S_{y},S_{z})$ represent the orbital and spin angular momentum operator, respectively. This has been studied by inclusion of heavy elements for damping in \ce{FePt_{1-x}Pd_x} \cite{he2013} and magnetostriction in Tb-doped FeGa \cite{niu2024}. However, practical usage of heavy elements is avoided considering their scarcity, high cost, and environmental damage. We recently observed a strong intrinsic link between magnetostriction and damping in \ce{(Fe_{1-x}Co_{x})_{4}N}, supported by theoretical calculations considering shifted Fermi level $E_\mathrm{F}$ \cite{ito2024}, as the calculated results for unshifted $E_\mathrm{F}$ shown in Fig. \ref{fig1}(a). The $\lambda_{100}$ corresponds to the magnetostriction due to tetragonal distortion, which empirically means a change in length along the [001] direction when the magnetization direction changes from the $ab$ plane to the $c$ direction. In this Letter, we demonstrate that there exists the intrinsic link between magnetostriction and magnetic damping not only for the \ce{(Fe_{1-x}Co_{x})_{4}N} but also for the \ce{Ni_{1-y}Co_{y}} from their electronic structure, facilitated by spin-conserving transitions, enabling simultaneous control of spatial and temporal responses under magnetic fields without significant SOI changes. These materials are also expected to undergo the sign change of magnetostriction by changing of $x$ ($y$) value, where a similar phenomenon is also found in the \ce{Ni_{x}Fe_{1-x}} \cite{endo2018}. Interestingly, our calculation shows that magnitude of damping seems affected by the sign of magnetostriction in [100] direction, where we may propose a possible scenario. This finding may improve the understanding the relation behind magnetostriction and magnetic damping.

First-principles density functional calculations were performed using VASP \cite{kresse1996} to study \ce{(Fe_{1-x}Co_{x})_{4}N} and \ce{Ni_{1-y}Co_{y}}, \textcolor{black}{each of which based on the} basic cubic structure of \ce{Fe_{4}N} and \ce{fcc-Ni}, respectively as illustrated in Fig. \ref{fig1}. \textcolor{black}{We employed the spin-polarized generalized gradient approximation (GGA) \cite{perdew1996} for the exchange-correlation energy and the projector augmented wave (PAW) \cite{blochl1994} method to accurately account for core-electrons effects. Starting from the cubic prototype, we optimized the lattice parameter by computing the total energy over a range of lattice constants and performing a spline interpolation to identify the equilibrium lattice constant, $c_0$, that minimizes the energy. From this optimized cubic geometry, we constructed the unstrained unit cells: a single-formula-unit cubic cell for \ce{(Fe_{1-x}Co_{x})_{4}N} and a two-formula-unit tetragonal cell for \ce{Ni_{1-y}Co_{y}} (rotated $45^\circ$ in-plane relative to the original cubic cell). Strained structures were generated by varying the out-of-plane lattice parameter \(c\); for each strain value $\varepsilon=(c-c_0)/c_0$, we fully relaxed the in-plane lattice parameters $a(=b)$, again using spline interpolation to locate the minimum-energy configuration. The \(k\)‑point meshes of \(30\times30\times30\) for \ce{(Fe_{1-x}Co_{x})_{4}N} and \(25\times25\times17\) for \ce{Ni_{1-y}Co_{y}} ensured convergence of the MAE, consistent with similar systems studied previously \cite{isogami2020,okabayashi2019}.} 

Without considering SOI, the face-centered sites (Fe2, Fe3, Fe4) exhibit identical atomic environments \cite{blanca2009}. For off-stoichiometric compositions between Fe-Co and Ni-Co, the virtual crystal approximation (VCA) \cite{bellaiche2000} was applied equivalently to all Fe sites \textcolor{black}{(both corner and face-centered)} in \ce{Fe_{4}N} and Ni sites, respectively. This assumption for \ce{Fe_{4}N} is warranted by previous theoretical \cite{monachesi2013} and experimental results \cite{ito2015}, indicating that Co has no preferential site in \ce{Fe_{4}N}. In addition, our choice to use VCA is justified by good agreement with experimental results for both of \ce{(Fe_{1-x}Co_{x})_{4}N} \cite{ito2024} and \ce{Ni_{1-y}Co_{y}} \cite{yamamoto1958}. \textcolor{black}{In our implementation, each "virtual" atom's potential, formal valence, augmentation charges, and non-local pseudopotential were mixed according to the target weighted composition, eliminating the need for large supercells since compositional effects are fully captured information within simple unit cell. Because the VCA is applied uniformly at all equivalent sites, inversion symmetry is preserved in this calculation.} 

Magnetostriction was calculated from the derivative of $E_\mathrm{001}$ and MAE with respect to $\varepsilon$ using the following equation \cite{wu1997,note1}:
\begin{equation}
\lambda_{100}=\dfrac{2}{3}\dfrac{d\mathrm{MAE}/d\varepsilon}{d^{2}E_\mathrm{001}/d\varepsilon^{2}}
\label{striction}
\end{equation}
The values of $E_\mathrm{001}$ and MAE were obtained by including the SOI in a self-consistent calculation for each different strained structure \textcolor{black}{(total energy SCF method)}. \textcolor{black}{We imposed a strict total-energy convergence criterion of \(10^{-7}\,\text{eV}\) and employed the tetrahedron method for Brillouin-zone integration}. Here, we define MAE as $\mathrm{MAE} = E_\mathrm{100} - E_\mathrm{001}$, where $E_\mathrm{100}$ and $E_\mathrm{001}$ represent the energies of magnetic materials with magnetization orientations in [100] and [001] directions, respectively. Furthermore, using the second-order perturbation analysis with respect to the SOI \cite{miura2022, note2}, we clarified the dominant atomic, orbital, and spin contributions to MAE hence also the $\textit{d}\mathrm{MAE}/d\varepsilon$. \textcolor{black}{This perturbative approach is particularly appropriate here because (1) it yields a nonzero MAE already at second-order under strain (whereas for the unstrained cubic lattice, nonzero contributions only appear at fourth order or higher), and (2) inversion symmetry remains intact thanks to the uniform implementation of the VCA, so no additional intraband terms are required (in contrast to low-symmetry cases \cite{cinal2022,cinal2024}. We also verified that MAE values from the perturbative analysis closely agree with those obtained via both the total-energy SCF method and the force-theorem methods, which can be used to understand the main contribution to the strain dependence of MAE.}
\begin{multline}
\mathrm{MAE}
= \mathrm{MAE}(\uparrow\Rightarrow\uparrow)+\mathrm{MAE}(\downarrow\Rightarrow\downarrow)\\
+\mathrm{MAE}(\uparrow\Rightarrow\downarrow)+\mathrm{MAE}(\downarrow\Rightarrow\uparrow)
\label{perturb}
\end{multline}
The magnetic damping was calculated using the torque correlation model \textcolor{black}{\cite{kambersky1976,gilmore2007}}, which is derived from noncollinear magnetic calculations including the SOI.
\begin{multline}
\alpha=\dfrac{g}{\pi M_{s}}\sum_{k}\sum_{nn'}|\Gamma^{-}_{nn'}(k)|^{2} \\
\times \dfrac{\delta}{(E_{\mathrm{F}}-\epsilon_{kn'\sigma})^{2}+\delta^{2}}\dfrac{\delta}{(E_{\mathrm{F}}-\epsilon_{kn\sigma})^{2}+\delta^{2}},
\label{damping}
\end{multline}
where the \textit{g} is electron's \textit{g}-factor, $M_{s}$ is the saturation magnetization, and $\Gamma^{-}_{nn'}(k)=\langle kn'\sigma'\vert [S_{-}, H_{\mathrm{SO}} ] \vert kn\sigma\rangle$, which is the matrix element of the spin-orbit torque operator $[S_{-}, H_{\mathrm{SO}} ]=\sum_{I}\xi_{I}(S_{-}L_{z}-S_{z}L_{-})$ between eigenstates including the SOI, with energies $\epsilon_{kn\sigma}$ labeled with wavevector \textit{k}, band index \textit{n}, and spin $\sigma$. These transitions occur near $E_\mathrm{F}$ and vanish after interaction with lattice, which are phenomenologically parameterized with scattering rate $\delta$. In this study, $\delta$ is estimated from the residual resistance of typical magnetic alloys (0.01 eV) \textcolor{black}{\cite{hiramatsu2022}}.

\begin{figure}
\includegraphics[width=\columnwidth]{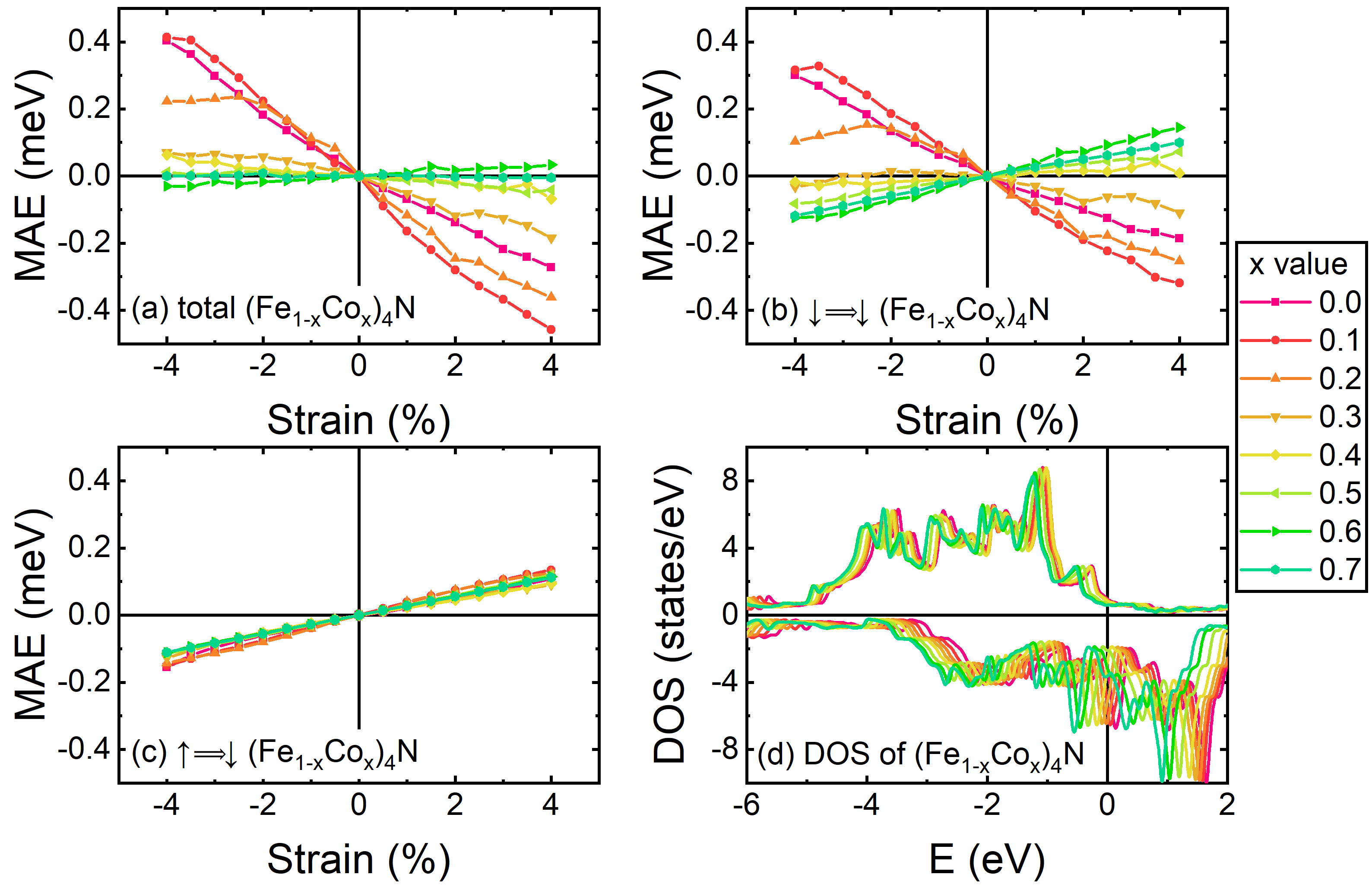}%
\caption{\label{fig2} Strain dependence of (a) total, (b) $\downarrow\Rightarrow\downarrow$, and (c) $\uparrow\Rightarrow\downarrow$ contribution to the MAE calculated from second-order perturbation analysis for \ce{(Fe_{1-x}Co_{x})_{4}N}. (d) Spin-resolved density of states of \ce{(Fe_{1-x}Co_{x})_{4}N}}
\end{figure}
\begin{figure}
\includegraphics[width=\columnwidth]{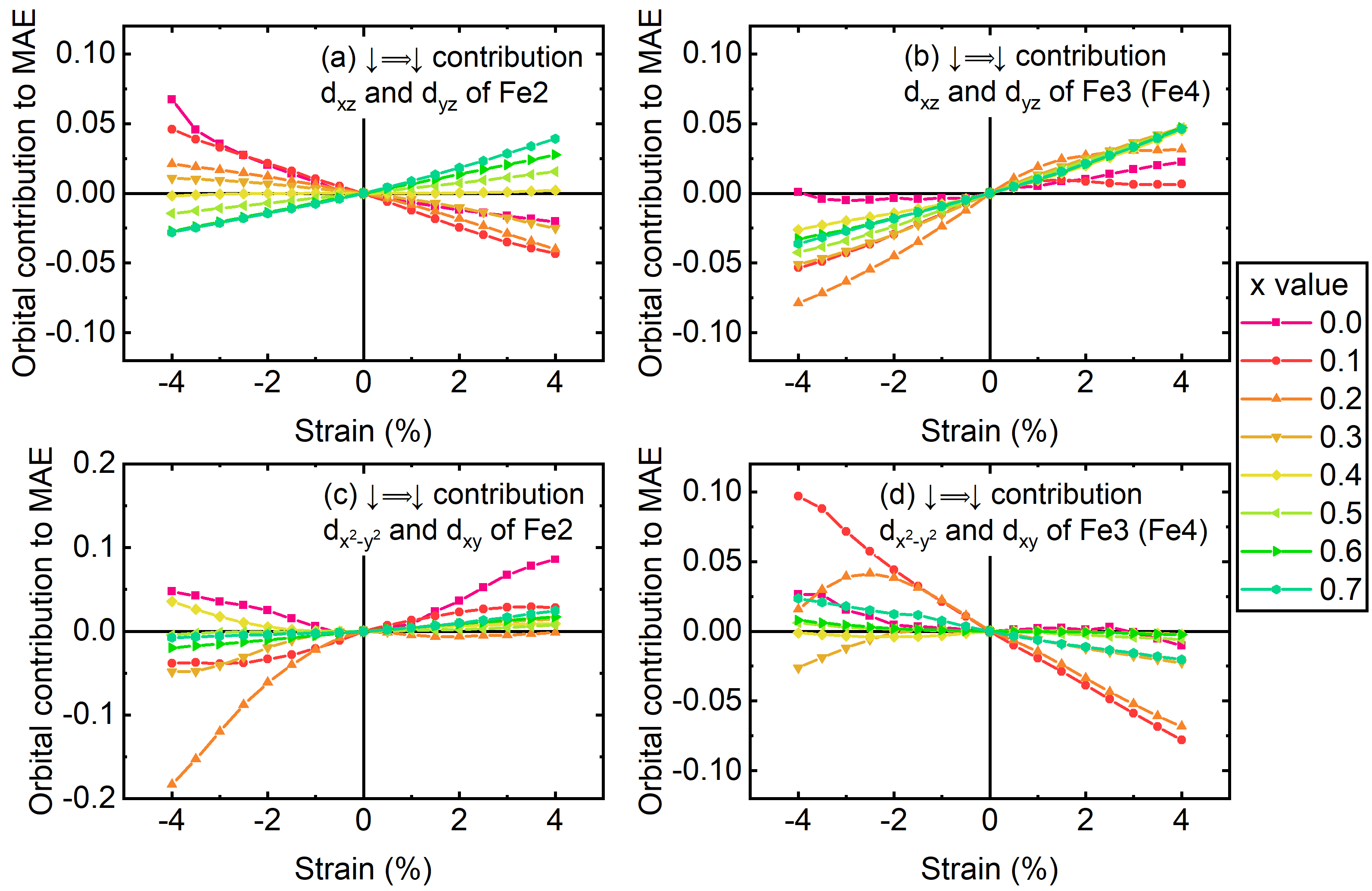}%
\caption{\label{fig3} Strain dependence of the $\downarrow\Rightarrow\downarrow$ normalized orbital contribution to the MAE of the \ce{(Fe_{1-x}Co_{x})_{4}N} calculated from second-order perturbation analysis for (a) $d_{xz}$ - $d_{yz}$ of Fe2, (b) $d_{xz}$ - $d_{yz}$ of Fe3 (Fe4) , (c) $d_{{x}^2-{y}^2}$ - $d_{xy}$ of Fe2, and (d) $d_{{x}^2-{y}^2}$ - $d_{xy}$ of Fe3 (Fe4) }
\end{figure}

The intrinsic relationship between magnetostriction and damping in both \ce{(Fe_{1-x}Co_{x})_{4}N} and \ce{Ni_{1-y}Co_{y}} is shown in Fig. \ref{fig1}. By changing the composition, we may simultaneously tune the magnetostriction and damping. In addition, one may notice that there are two composition regions: a region that has negative magnetostriction, which also exhibits large magnetic damping, and regions with positive magnetostriction that are associated with small magnetic damping. Although one data point from \ce{Fe_{4}N} deviates from this general trend, most data points cluster within the described regions. To explore this relationship further, we focus on \ce{(Fe_{1-x}Co_{x})_{4}N} first. The second-order perturbation analysis reveals the significant contributions to $d\mathrm{MAE}/d\varepsilon$ for \ce{(Fe_{1-x}Co_{x})_{4}N} [Figs. \ref{fig2}(a)-(c)]. As shown in Fig. \ref{fig2}(a), the gradient of total $d\mathrm{MAE}/d\varepsilon$ corresponds to the sign and magnitude of the magnetostriction coefficient. From the second-order perturbation analysis of MAE \cite{miura2022}, it is found that the spin-conserving $\downarrow\Rightarrow\downarrow$ [Fig. \ref{fig2}(b)] and spin-flip $\uparrow\Rightarrow\downarrow$ [Fig. \ref{fig2}(c)] contributions dominate the matrix elements near $E_\mathrm{F}$. In order to understand the role of each contribution, let us focus on three compositional regions. First, at low Co contents ($x < 0.3$), the spin-conserving process is more dominant than the competing spin-flip term, which controls the behavior of the total MAE. At moderate level of Co content ($0.3 \leq x \leq 0.5$), $d\mathrm{MAE}(\downarrow\Rightarrow\downarrow)/d\varepsilon$ become smaller compared to the positive $d\mathrm{MAE}(\uparrow\Rightarrow\downarrow)/d\varepsilon$ leading to a sign change in $\lambda_{100}$. When Co content is further increased ($x > 0.5$), both of $d\mathrm{MAE}(\downarrow\Rightarrow\downarrow)/d\varepsilon$ and $d\mathrm{MAE}(\uparrow\Rightarrow\downarrow)/d\varepsilon$ are positive. The negligible contributions of spin-conserving $\uparrow\Rightarrow\uparrow$ and spin-flip $\downarrow\Rightarrow\uparrow$ are not shown. This behavior can be understood from the electronic structure of \ce{(Fe_{1-x}Co_{x})_{4}N} with more than half-filled transition metal elements, where the majority-spin states are fully occupied and the electronic states around $E_\mathrm{F}$ mainly come from the minority-spin states [Fig. \ref{fig2}(d)]. 

The relationship between magnetic damping and magnetostriction can be explained as follows. An intrinsic magnetic damping is described using the torque correlation model proposed by Kambersky \cite{kambersky1976}, as expressed in Eq. (\ref{damping}). In general, the precession motion of magnetization is damped by interactions with conduction electrons through the SOI. This interaction is mediated by spin-orbit torque, which contains operators for spin-flip ($S_{-}L_{z}$) and spin-conserving ($S_{z}L_{-}$) contributions. Note that $\vert kn\sigma\rangle = \Sigma_{j\mu} c^{kn}_{j\mu \sigma} \vert\mu\sigma\rangle e^{ikR_{j}}$, where the Bloch states $\vert kn\sigma\rangle$ are expanded with an orthogonal basis of atomic orbitals labeled $\mu$ (or $\nu$) for site \textit{j} with atomic position $R_{j}$ in the unit cell and the coefficient of the expansion $c^{kn}_{j\mu \sigma}$ from projected state on each localized atomic orbitals. The precession is annihilated and resulting in the creation of the electron-hole pairs which may occupy the same (intraband) or different (interband) band indices $n, n'$. This process concludes with scattering between electron-hole pairs and the lattice, parameterized by the electron-lattice scattering rate $\delta$ into an equilibrium state. It is known that intraband terms are proportional to the density of states at $E_\mathrm{F}$ \textcolor{black}{\cite{gilmore2008}}. Assuming a pure spin state at $E_\mathrm{F}$, spin-flip contributions may be neglected in the intraband mechanism, leaving the spin-conserving matrix elements:
\begin{equation}
\Gamma^{-}_{nn'}(k)=\sum\limits_{I}\xi_{I}\sum\limits_{\mu\mu'}c^{kn'*}_{I\mu'\downarrow}c^{kn}_{I\mu\downarrow}\langle \mu'\vert L_{-}\vert \mu\rangle
\end{equation}
Meanwhile, the spin-conserving term in MAE can also be expressed as follow:
\begin{multline}
\mathrm{MAE}(\downarrow\Rightarrow\downarrow)=\sum_{II'}\xi_{I}\xi_{I'}\sum_{\nu\nu'\mu\mu'} [\langle \nu\vert L_{z} \vert \nu'\rangle \langle \mu'\vert L_{z} \vert \mu\rangle -  \\
\langle \nu\vert L_{x} \vert \nu'\rangle \langle \mu'\vert L_{x} \vert \mu\rangle ] G^{\downarrow\downarrow}_{II'}(\nu\nu';\mu\mu')
\label{maedd}
\end{multline}
where the 
\begin{equation}
G^{\downarrow\downarrow}_{II'}(\nu\nu';\mu\mu')=\sum_{k}\sum^{\mathrm{occ}}_{n}\sum^{\mathrm{unocc}}_{n'} \dfrac{c^{kn*}_{I'\nu\downarrow}c^{kn'}_{I'\nu'\downarrow}c^{kn'*}_{\textcolor{black}{I}\mu'\downarrow}c^{kn}_{\textcolor{black}{I}\mu\downarrow}}{\epsilon_{kn'\sigma'}-\epsilon_{kn\sigma}} 
\label{jldos}
\end{equation}
Thus, spin-polarized magnetic materials with dominant minority-spin states exhibit nonzero $c^{kn}_{I\mu\downarrow}$ values, giving nonzero spin-conserving transition matrix elements of the angular momentum operator responsible for both damping ($\langle \mu'\vert L_{-}\vert \mu\rangle$) and magnetoelasticity ($\langle \mu'\vert L_{z}\vert \mu\rangle$, $\langle \mu'\vert L_{x}\vert \mu\rangle$). This connection facilitates the strong correlation between magnetostriction and magnetic damping. Additionally, one can propose an intuitive picture where negative magnetostriction is associated with relatively large magnetic damping. In all the systems, the increase of the strain from negative to positive increases the magnetization as shown in Fig. S1 in Supplemental Materials \textcolor{black}{\cite{supplem}}. This means that the increase of the strain increases the exchange splitting of the system. In other words, the increase of the strain shifts the majority-spin and minority-spin states to the lower and higher energy sides, respectively. These effects couple the strain and the magnetic damping through the density of states. When $E_\mathrm{F}$ is located slightly to the left of a peak in the minority spin states (for small $x$ and $y$ values in \ce{(Fe_{1-x}Co_{x})_{4}N} and \ce{Ni_{1-y}Co_{y}} cases, respectively), it results in large damping due to the proximity to the peak of the minority spin states. At the same time, there is a negative strain dependence of the density of states because strain pushes the peak farther from $E_\mathrm{F}$, thereby reducing the states at $E_\mathrm{F}$. This leads to negative magnetoelasticity and magnetostriction, because of the reduction of the spin-conserving term in the MAE with increasing the strain. Conversely, when $E_\mathrm{F}$ is situated in the valley of the minority spin states (for large $x$ and $y$ values in \ce{(Fe_{1-x}Co_{x})_{4}N} and \ce{Ni_{1-y}Co_{y}} cases, respectively), the system exhibits small damping. In this scenario, strain enhances the states at $E_\mathrm{F}$ by bringing them closer to the nearest peak of occupied minority spins, leading to positive magnetostriction.

In addition to the relationship between magnetostriction and damping, the substitution of Fe by Co results in a sign change in magnetostriction in \ce{(Fe_{1-x}Co_{x})_{4}N}, as shown in Fig. \ref{fig1}(a). We now examine the role of atomic sites and orbitals in this sign change of the $\downarrow\Rightarrow\downarrow$ magnetoelasticity of \ce{(Fe_{1-x}Co_{x})_{4}N}, as shown in Fig. \ref{fig3}. It was observed that the face-centered atoms (Fe2, Fe3, Fe4) significantly influence the strain dependence of the MAE, compared to the corner site (Fe1) (not shown). Notably, the dominant orbital contribution in \ce{(Fe_{1-x}Co_{x})_{4}N} varies depending on the symmetry of the atomic sites when the magnetization direction is changed. For Fe2 (aligned along the \textit{c}-axis), the MAE contribution from the transition between $d_{xz}$ and $d_{yz}$ strongly correlates with the compositional dependence of the magnetoelasticity [Fig. \ref{fig3}(a)]. This substantial contribution to MAE persists because both orbitals remain degenerate even under tetragonal distortion [see Fig. S2(b) in Supplemental Materials \textcolor{black}{\cite{supplem}}]. Furthermore, for Fe3 (aligned along the \textit{b}-axis) and Fe4 (aligned along the \textit{a}-axis), the $d_{{x}^2-{y}^2}$ and $d_{xy}$ orbitals dictate the magnetoelastic behavior [Fig. \ref{fig3}(d)]. These $d_{{x}^2-{y}^2}$ and $d_{xy}$ orbitals do not remain degenerate under tetragonal distortion at the same atom, but their similar behavior in Fe3-$d_{{x}^2-{y}^2}$ (Fe4-$d_{{x}^2-{y}^2}$) and Fe4-$d_{xy}$ (Fe3-$d_{xy}$) can be explained by the symmetry invariance of the $d_{{x}^2-{y}^2}$ and $d_{xy}$ orbitals between Fe3 and Fe4 sites [see Fig. S2(c)-(d) in Supplemental Materials \textcolor{black}{\cite{supplem}}]. It is important to note that similar contributions were not observed for $d_{xz}$ and $d_{yz}$ in Fe3 or Fe4 [Fig. \ref{fig3}(b)], nor for $d_{{x}^2-{y}^2}$ and $d_{xy}$ in Fe2 [Fig. \ref{fig3}(c)], nor any other orbitals [see Fig. S3 in Supplemental Materials \textcolor{black}{\cite{supplem}}], highlighting the role of the symmetry in the orbital contribution to magnetoelasticity. Note that the orbital contribution is defined as the sum of the $|\langle \mu'\vert L_{x(z)} \vert \mu\rangle|^{2} G^{\downarrow\downarrow}_{II'}(\mu'\mu';\mu\mu) + |\langle \mu\vert L_{x(z)} \vert \mu'\rangle|^{2} G^{\downarrow\downarrow}_{II'}(\mu\mu;\mu'\mu')$, normalized so that the zero value for the unstrained structure ($\varepsilon=0$) was obtained.

\begin{figure}
\includegraphics[width=\columnwidth]{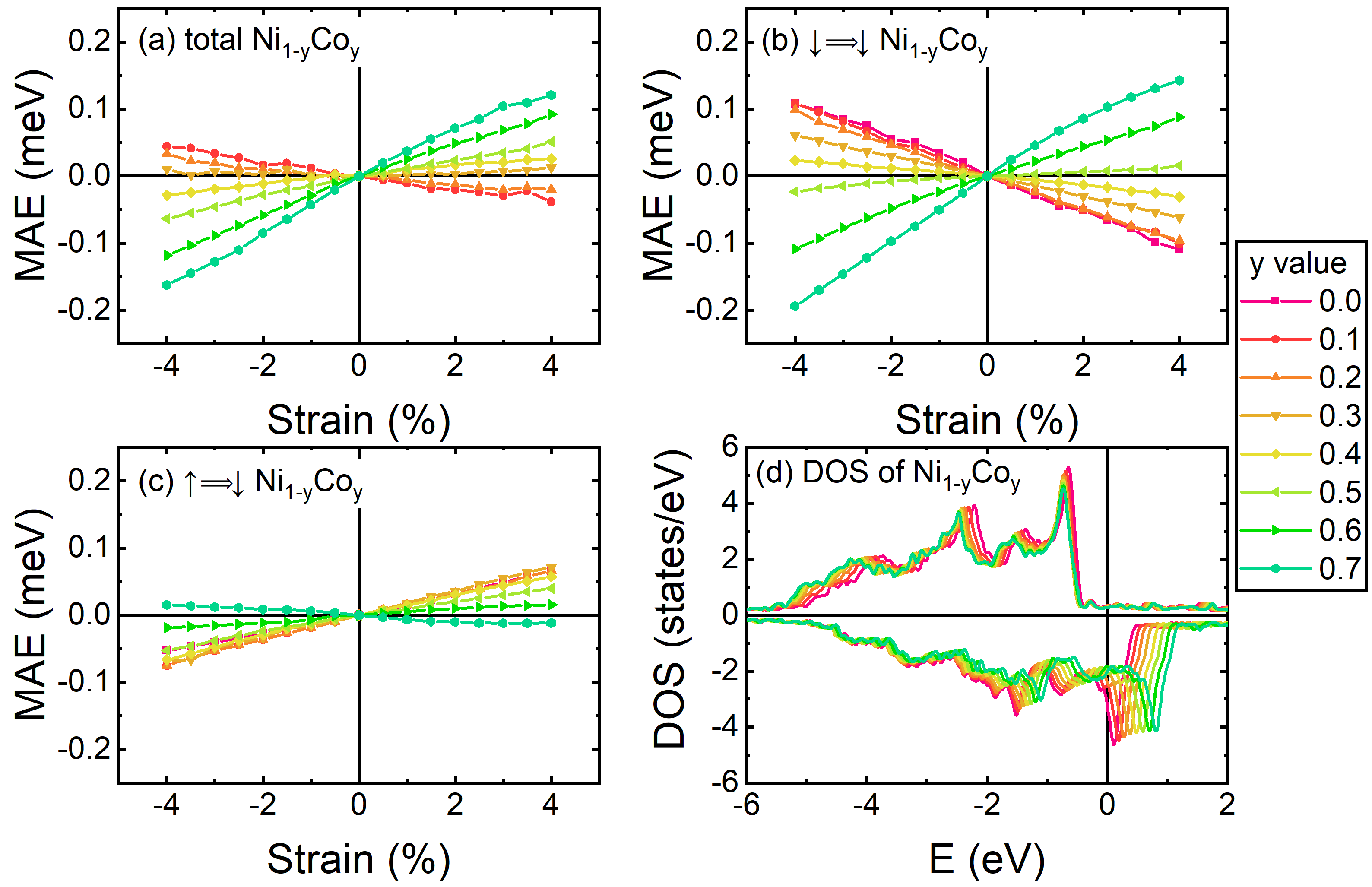}%
\caption{\label{fig4} Strain dependence of (a) total, (b) $\downarrow\Rightarrow\downarrow$, and (c) $\uparrow\Rightarrow\downarrow$ contribution to the MAE calculated from second-order perturbation analysis for \ce{Ni_{1-y}Co_{y}}. (d) Spin-resolved density of states of \ce{fcc-Ni_{1-y}Co_{y}}}
\end{figure}
\begin{figure}
\includegraphics[width=\columnwidth]{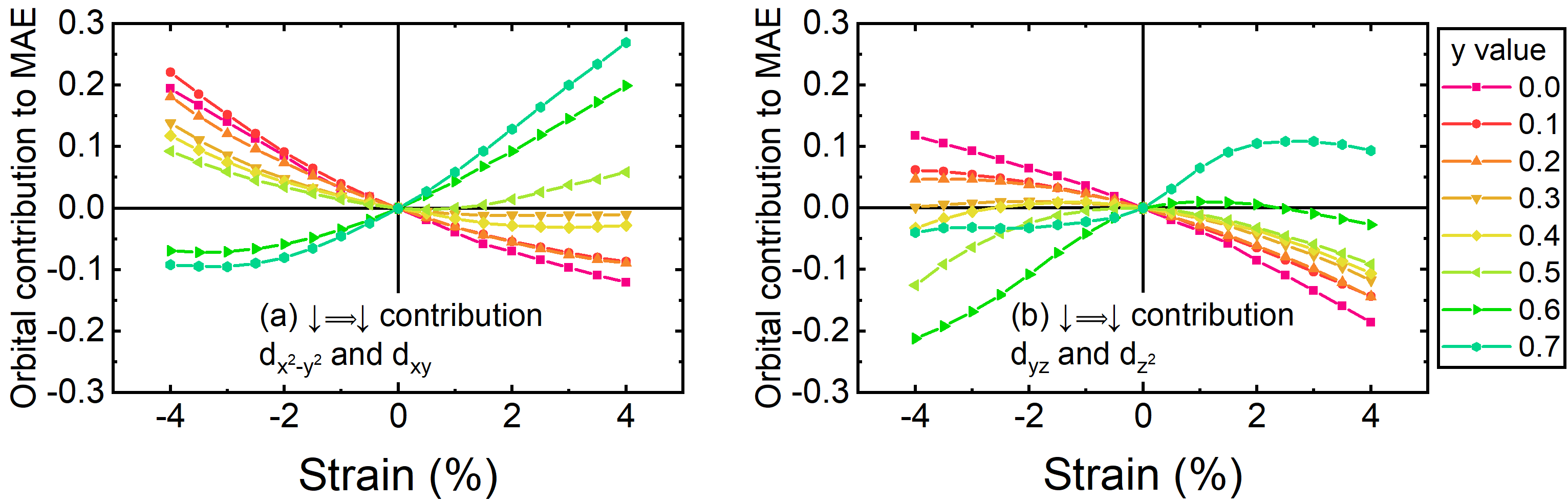}%
\caption{\label{fig5} Strain dependence of the $\downarrow\Rightarrow\downarrow$ normalized orbital contribution to the MAE calculated from second-order perturbation analysis for (a) $d_{{x}^2-{y}^2}$ - $d_{xy}$ and (b) $d_{yz}$ - $d_{{z}^2}$ of \ce{Ni_{1-y}Co_{y}} }
\end{figure}

In the case of \ce{Ni_{1-y}Co_{y}}, the main contribution to the sign change of magnetostriction is also expected to be the same as that of \ce{(Fe_{1-x}Co_{x})_{4}N} [Figs. \ref{fig4}(a)-(c)]. The total $d\mathrm{MAE}/d\varepsilon$ [Fig. \ref{fig4}(a)] is still dominated by the spin-conserving term $d\mathrm{MAE}(\downarrow\Rightarrow\downarrow)/d\varepsilon$ [Fig. \ref{fig4}(b)]. The only notable difference with \ce{(Fe_{1-x}Co_{x})_{4}N} is that, with the inclusion of Co ($y > 0.6$), the spin-flip term $d\mathrm{MAE}(\uparrow\Rightarrow\downarrow)/d\varepsilon$ slightly shifts from positive to a small negative value [Fig. \ref{fig4}(c)], while the spin-flip term $d\mathrm{MAE}(\uparrow\Rightarrow\downarrow)/d\varepsilon$ of \ce{(Fe_{1-x}Co_{x})_{4}N} is always positive [Fig. \ref{fig2}(c)]. Since the total density of states at $E_\mathrm{F}$ in this system is mainly from minority-spin state [Fig. \ref{fig4}(d)], the same mechanism of spin-conserving transitions mediated by minority-spin governs the MAE and its strain dependence, while also controlling the intraband transitions in the minority-spin states for the magnetic damping, due to the absence of majority-spin \textit{d} states around $E_\mathrm{F}$.

In \ce{Ni_{1-y}Co_{y}}, two significant transitions $d_{{x}^2-{y}^2}$-$d_{xy}$ and $d_{yz}$-$d_{{z}^2}$ determine the compositional dependence of $d\mathrm{MAE}/d\varepsilon$ [Figs. \ref{fig5}(a)-(b)]. Both orbital contributions show the decrease of MAE with increasing the strain in fcc-Ni, consistent with the negative magnetostriction of fcc-Ni observed in many previous experiments. The substitution of Ni by Co significantly affects these two transitions, leading to the increase of MAE with increasing the strain. It is worth noting that Co doping does not significantly affect other orbital contributions to the MAE and its strain dependence [see Fig. S4 in the Supplemental Materials \textcolor{black}{\cite{supplem}}]. Unlike the Fe sites in \ce{(Fe_{1-x}Co_{x})_{4}N}, the Ni sites in \ce{Ni_{1-y}Co_{y}} have the same atomic environment as each other and equally contribute to the magnetoelasticity. These results emphasize the important role of crystal symmetry and the role of light elements (here, Nitrogen atom) in understanding the effects of atomic substitution on magnetoelasticity.

In summary, we theoretically investigated the correlation between magnetostriction and intrinsic magnetic damping as a function of composition in \ce{(Fe_{1-x}Co_{x})_{4}N} and \ce{Ni_{1-y}Co_{y}} alloys. We found that damping is larger when the magnetostriction constant is negative and smaller when the magnetostriction constant is positive. This is because the magnitude of magnetization changes (exchange splitting changes) when strain is applied, and the change in exchange splitting shifts the density of states near the Fermi level, affecting both magnetostriction and magnetic damping. We also found that the presence of locally degenerate orbitals is important with respect to the magnetostriction constant and that the presence of degenerate orbitals in the same atom or atoms of the same symmetry dominantly contributes to the strain dependence of the MAE. These results suggest the possibility of controlling magnetization dynamics by the application of strain to magnetic materials and provide important guidelines for future experiments.

\begin{acknowledgments}
The authors are grateful to T. Tadano and G. Xing of NIMS for their valuable discussions on this paper. This research was partially supported by Grants-in-Aid for Scientific Research (Grant No. JP22H04966) from the Japan Society for the Promotion of Science and the Japan Science and Technology Agency (JST) CREST (Grant No. JPMJCR21O1). The calculations were performed on the Numerical Materials Simulator at NIMS.
\end{acknowledgments}


\begin{thebibliography}{99}
\bibitem{joule1842} J. P. Joule, \textcolor{black}{On a new class of magnetic forces}, Annals Electric. Magn. Chem. \textbf{8}, 219 (1842).
\bibitem{bidwell1890} S. Bidwell, \textcolor{black}{On the effect of tension upon magnetic changes of length in wires of iron, nickel, and cobalt}, Proc. Roy. Soc. \textbf{47}, 469 (1890).
\bibitem{kittel1949} C. Kittel, \textcolor{black}{Physical theory of ferromagnetic domains}, Rev. Mod. Phys. \textbf{21}, 541 (1949).
\bibitem{wu1997} R. Wu, L. Chen, and A. Freeman, \textcolor{black}{First principles determination of magnetostriction in bulk transition metals and thin films}, J. Magn. Magn. Mater. \textbf{170}, 103 (1997).
\bibitem{clark1972} A. E. Clark, and H. S. Belson, \textcolor{black}{Giant room-temperature magnetostrictions in \ce{TbFe2} and \ce{DyFe2}}, Phys. Rev. B \textbf{5}, 3642 (1972).
\bibitem{suhl1998} H. Suhl, \textcolor{black}{Theory of the magnetic damping constant}, IEEE Trans. Magn. \textbf{34}, 1834 (1998).
\bibitem{vittoria2010} C. Vittoria, S. D. Yoon, and A. Widom, \textcolor{black}{Relaxation mechanism for ordered magnetic materials}, Phys. Rev. B \textbf{81}, 014412 (2010).
\bibitem{rossi2005} E. Rossi, O. G. Heinonen, and A. H. MacDonald,  \textcolor{black}{Dynamics of magnetization coupled to a thermal bath of elastic modes}, Phys. Rev. B \textbf{72}, 174412 (2005).
\bibitem{chumak2021} O. M. Chumak, A. Pacewicz, A. Lynnyk, B. Salski, T. Yamamoto, T. Seki, J. Z. Domagala, H. Glowi\'{n}ski, K. Takanashi, L. T. Baczewski, H. Szymczak, and A. Nabialek, \textcolor{black}{Magnetoelastic interactions and magnetic damping in \ce{Co2Fe_{0.4}Mn_{0.6}Si} and \ce{Co2FeGa_{0.5}Ge_{0.5}} Heusler alloys thin films for spintronic applications}, Sci. Rep. \textbf{11}, 7608 (2021).
\bibitem{endo2018} Y. Endo, O. Mori, Y. Shimada, S. Yabukami, S. Sato, and R. Utsumi, \textcolor{black}{Study on measurement technique for magnetization dynamics of thin films}, Appl. Phys. Lett. \textbf{112}, 252403 (2018).
\bibitem{kambersky1976} V. Kambersk\'{y}, \textcolor{black}{On ferromagnetic resonance damping in metals}, Czech. J. Phys. \textbf{26}, 1366 (1976).
\bibitem{ota2018} S. Ota, A. Ando, and D. Chiba, \textcolor{black}{A flexible giant magnetoresistive device for sensing strain direction}, Nat. Electron. \textbf{1}, 124 (2018).
\bibitem{he2013} P. He, X. Ma, J. W. Zhang, H. B. Zhao, G. L\"{u}pke, Z. Shi, and S. M. Zhou, \textcolor{black}{Quadratic scaling of intrinsic Gilbert damping with spin-orbital coupling in $L1_{0}$ FePdPt films: experiments and \textit{ab initio} calculation}, Phys. Rev. Lett. \textbf{110}, 077203 (2013).
\bibitem{niu2024} J. Niu, K. Yan, Y. Xu, Y. Wu, E. Liu, Z. Fu, J. Li, X. Gao, X. Mu, B. Liu, X. Wang, Y. Li, J. Wang, and C. Jiang, \textcolor{black}{Understanding the intrinsic mechanism of the giant magnetostriction in binary and alloyed FeGa solid solutions}, Phys. Rev. B \textbf{109}, 014417 (2024).
\bibitem{ito2024} K. Ito, I. Kurniawan, Y. Shimada, Y. Miura, Y. Endo, and T. Seki, \textcolor{black}{Giant tunability of magnetoelasticity in \ce{Fe4N} system as a platform to unveil correlation between magnetostriction and magnetic damping, Commun. Mater. \textbf{6}, 53 (2025).}
\bibitem{kresse1996} G. Kresse and J. Furthm\"uller , \textcolor{black}{Efficient iterative schemes for \textit{ab initio} total-energy calculations using a plane-wave basis set}, Phys. Rev. B \textbf{54}, 11169 (1996).
\bibitem[\textcolor{black}{17}]{perdew1996} \textcolor{black}{J. P. Perdew, K. Burke, and M. Ernzerhof, Generalized gradient approximation made simple, Phys. Rev. Lett. \textbf{77}, 3865 (1996).}
\bibitem[\textcolor{black}{18}]{blochl1994} \textcolor{black}{P. E. Bl\"ochl, Projector augmented-wave method, Phys. Rev. B \textbf{50}, 17953 (1994).}
\bibitem[\textcolor{black}{19}]{isogami2020} \textcolor{black}{S. Isogami, K. Masuda, and Y. Miura, Contributions of magnetic structure and nitrogen to perpendicular magnetocrystalline anisotropy in antiperovskite $\varepsilon$-\ce{Mn4N}, Phys. Rev. Mater \textbf{4}, 014406 (2020).}
\bibitem[\textcolor{black}{20}]{okabayashi2019} 
\textcolor{black}{J. Okabayashi, Y. Miura, and T. Taniyama, Strain-induced reversible manipulation of orbital magnetic moments in Ni/Cu multilayers on ferroelectric \ce{BaTiO3}, npj Quantum Mater. \textbf{4}, 21 (2019).}
\bibitem[\textcolor{black}{21}]{blanca2009} E. L. P. y. Blanc\'{a}, J. Desimoni, N. E. Christensen, H. Emmerich, and S. Cottenier, \textcolor{black}{The magnetization of $\gamma'$-\ce{Fe4N}: theory vs. experiment}, Phys. Status Solidi B \textbf{246}, 909 (2009).
\bibitem[\textcolor{black}{22}]{bellaiche2000} L. Bellaiche and D. Vanderbilt, \textcolor{black}{Virtual crystal approximation revisited: Application to dielectric and piezoelectric properties of perovskites}, Phys. Rev. B \textbf{61}, 7877
(2000).
\bibitem[\textcolor{black}{23}]{monachesi2013} P. Monachesi, T. Bj\"{o}rkman, T. Gasche, and O. Eriksson, \textcolor{black}{Electronic structure and magnetic properties of Mn, Co, and Ni substitution of Fe in \ce{Fe4N}}, Phys. Rev. B \textbf{88}, 054420 (2013).
\bibitem[\textcolor{black}{24}]{ito2015} K. Ito, T. Sanai, Y. Yasutomi, T. Gushi, K. Toko, H. Yanagihara, M. Tsunoda, E. Kita, and T. Suemasu, \textcolor{black}{M{\"o}ssbauer study on epitaxial \ce{Co_{x}Fe_{4-x}N} films grown by molecular beam epitaxy}, J. Appl. Phys. \textbf{117}, 17B717 (2015).
\bibitem[\textcolor{black}{25}]{yamamoto1958} M. Yamamoto, T. Nakamichi, \textcolor{black}{Magnetostriction constants of nickel-copper and nickel-cobalt alloys}, J. Phys. Soc. Japan \textbf{13}, 228 (1958).
\bibitem[\textcolor{black}{26}]{note1} The strain dependence of $E_{001}$ and MAE is fitted using interpolation.
\bibitem[\textcolor{black}{27}]{miura2022} Y. Miura and J. Okabayashi, \textcolor{black}{Understanding magnetocrystalline anisotropy based on orbital and quadrupole moments}, J. Phys.: Condens. Matter \textbf{34}, 473001 (2022).
\bibitem[\textcolor{black}{28}]{note2} The values of $\xi$ used for Fe, Co, Ni, and N in the second-order perturbation calculation are 54.3, 69.4, 87.2, and 20 meV, respectively. \textcolor{black}{The coefficient of the expansion from projected state on each localized atomic orbitals were obtained by simple projection with option LORBIT=2 in VASP.}
\bibitem[\textcolor{black}{29}]{cinal2022} \textcolor{black}{M. Cinal, Magnetic anisotropy and orbital magnetic moment in Co films and Co/X bilayers (X=Pd and Pt), Phys. Rev. B \textbf{105}, 104403 (2022)}
\bibitem[\textcolor{black}{30}]{cinal2024} \textcolor{black}{M. Cinal, Intraband terms of magnetocrystalline anisotropy energy in layered systems without inversion symmetry, Phys. Rev. B \textbf{109}, 024424, (2024).}
\bibitem[\textcolor{black}{31}]{gilmore2007} K. Gilmore, Y. U. Idzerda, and M. D. Stiles, \textcolor{black}{Identification of the dominant precession-damping mechanism in Fe, Co, and Ni by first-principles calculations}, Phys. Rev. Lett. \textbf{99}, 027204 (2007).
\bibitem[\textcolor{black}{32}]{hiramatsu2022} R. Hiramatsu, D. Miura, and A. Sakuma, \textcolor{black}{First-principles calculations for Gilbert damping constant at finite temperature}, Appl. Phys. Express \textbf{15}, 013003 (2022).
\bibitem[\textcolor{black}{33}]{gilmore2008} K. Gilmore, Y. U. Idzerda, and M. D. Stiles, \textcolor{black}{Spin-orbit precession damping in transition metal ferromagnets}, J. Appl. Phys. \textbf{103}, 07D303 (2008).
\bibitem[\textcolor{black}{34}]{supplem} See Supplemental Material at [URL will be inserted by publisher] for strain dependence of magnetization and orbital contribution to MAE of \ce{(Fe_{1-x}Co_{x})_{4}N} and \ce{Ni_{1-y}Co_{y}}, and atomic-resolved partial density of states of tetragonally distorted \ce{Fe_{4}N}.
\end{thebibliography}

\end{document}